\documentclass[12pt,preprint]{aastex}

\newcommand{\ltappeq}{\raisebox{-0.6ex}{$\,\stackrel
{\raisebox{-.2ex}{$\textstyle <$}}{\sim}\,$}}

\slugcomment{accepted for publication in ApJ}

\shorttitle{A Blue Straggler with an Active Companion in 47 Tuc}

\shortauthors{Knigge et al.}

\begin{document}



\title{A Blue Straggler Binary with Three Progenitors in the Core of a Globular Cluster?$^1$} 

\footnotetext[1]{Based on observations with the NASA/ESA Hubble Space
Telescope, obtained at the Space Telescope Science Institute, which is
operated by the Association of Universities for Research in Astronomy,
Inc. under NASA contract No. NAS5-26555.}


\author{C. Knigge\altaffilmark{1}, R. L. Gilliland\altaffilmark{2}, 
A. Dieball\altaffilmark{1}, D. R. Zurek\altaffilmark{3},
M. M. Shara\altaffilmark{3}, K. S. Long\altaffilmark{2}}

\altaffiltext{1}{School of Physics \& Astronomy, University of
Southampton, Highfield, Southampton SO17 1BJ, UK} 

\altaffiltext{2}{Space Telescope Science Institute, 3700 San Martin
Drive, Baltimore, MD 21218, USA}

\altaffiltext{3}{Department of Astrophysics, American Museum of
Natural History, New York, NY  10024}

\email{christian@astro.soton.ac.uk, gillil@stsci.edu,
andrea@astro.soton.ac.uk, zurek@amnh.org, mshara@amnh.org,
long@stsci.edu} 

\begin{abstract}

We show that the X-ray source W31 in the core of the globular cluster 
47~Tucanae is physically associated with the bright blue straggler
BSS-7. The two sources are astrometrically matched to 0.061\arcsec,
with a chance coincidence  probability of less than 1\%. We then
analyse optical time-series photometry obtained with the {\em Hubble 
Space Telescope} (HST) and find that BSS-7 displays a 1.56~day 
periodic signal in the I band. We also construct a broad-band
(far-ultraviolet through far-red) spectral energy distribution for
BSS-7 and fit this with single and binary models. The binary model
is a better fit to the data, and we derive the corresponding stellar
parameters. 

All of our findings are consistent with BSS-7 being a detached binary
consisting of a blue straggler primary with an X-ray-active,
upper-main-sequence companion. The formation of such a system would
necessarily involve at least three stars, which is consistent with
recent N-body models in which blue stragglers often form via multiple 
encounters that can involve both single and binary stars. However, we
cannot yet entirely rule out the possibility that BSS-7 descended
directly from a binary system via mass transfer. The system parameters
needed to distinguish definitively between these scenarios may be
obtainable from time-resolved spectroscopy.

\end{abstract}


\section{Introduction}
\label{intro}

Blue straggler stars (BSSs) are objects with roughly main-sequence
colors and magnitudes that are significantly bluer and brighter than the
main-sequence turn-off (MSTO) point appropriate to their
environment. Their name derives from the fact that they should have
evolved off the MS long ago, i.e. they appear to ``straggle'' on the 
evolutionary path taken by ordinary stars. Since the MSTO is a
function of age, BSSs are most easily detected in 
co-eval stellar populations, such as star clusters. Globular clusters
(GCs) provide a particularly interesting setting for studying BSSs,
since the ultra-high stellar densities found in the cores of GCs (up
to one million stars per cubic parsec) open up new channels for BSS
formation via dynamical encounters between cluster members (e.g. Hut
et al. 1992).

BSS have been found in essentially all galactic globular clusters
(GCs) (e.g. Piotto et al. 2004). The most popular scenarios for BSS
formation in GCs are (i) direct stellar collisions between cluster
members; (ii) mass transfer in and/or coalescence of primordial binary
systems. Both mechanisms are likely to contribute significantly to the
observed BSS populations in GCs (Davies et al. 2004), but dynamical
interactions are likely to dominate BSS formation in dense cluster
cores (e.g. Ferraro et al. 2004; Mapelli et al. 2004). 

However, recent state-of-the-art numerical simulations suggest that
even this picture is probably too simplistic. In particular, these
models indicate that BSSs found in dense environments may have led
much more promiscuous lives than a simple one-time collision between 
two single stars. Instead, many of these objects may have undergone
multiple dynamical interactions involving both single and binary stars 
(Hurley \& Shara 2002; Hurley et al. 2005).

Here, we present astrometric and photometric evidence that a
previously known BSS in the core of the cluster 47~Tucanae is
actually a close binary system with a BSS primary and an active lower
mass secondary. If the binary is detached and the secondary is on the
main seqeuence, this system cannot have been formed simply from a
primordial binary or from just a single 2-body encounter. Its
existence would therefore imply the presence of primordial triple
systems in GC cores, or, more probably, the importance of more complex
dynamical encounters for BSS formation. However, we cannot yet
completely rule out the possibility that BSS-7 descended directly from
a binary via mass transfer. A radial velocity study will be needed to
distinguish clearly between these possibilities.

\section{Observations \& Analysis}

The blue straggler BSS-7 in 47 Tucanae was originally discovered in
ultraviolet images of the cluster core obtained with the Faint Object
Camera onboard the {\em Hubble Space Telescope} (HST) (Paresce et
al. 1991). The work on this system described here was prompted by a recently
completed effort to astrometrically link a set of deep,
far-ultraviolet (FUV) HST observations of 47 Tuc (Knigge et al. 2002
[hereafter K02]) with the deep X-ray images of the cluster that have
been obtained with the Chandra X-ray Observatory (Grindlay et al. 2001
[hereafter G01]; Heinke et al. 2005 [hereafter H05]). 
In the process, we discovered a tight match between BSS-7 and the
Chandra source W31, which was detected at L$_X$(0.5-6.0 keV) =
$4.8^{+0.7}_{-0.5} \times 10^{30}$~erg~s$^{-1}$\ in the 2002
ultra-deep Chandra observations of 47 Tuc and which displayed X-ray
variability on time scales of hours and days (H05). 

\subsection{Astrometry}
\label{astrometry}

Full details of our astrometric analysis will be provided in a
separate publication. Briefly, the FUV imaging observations were
obtained with the STIS/FUV-MAMA detector onboard HST. The F25QTZ
filter was used, giving a FUV bandpass with a mean wavelength of
1598~\AA. The STIS/FUV-MAMA field of view is about 25\arcsec
$\times$~25\arcsec, and is sampled 
at roughly 0.025\arcsec~pix$^{-1}$. We have also registered a deep
optical image obtained with HST onto the FUV reference frame. This
image was constructed from exposures with  the WFPC2/PC/F336W
instrument/detector/filter combination, which provides a bandpass with
mean wavelength 3367~\AA. The native platescale of the WFPC2/PC data
is about 0.045\arcsec~pix$^{-1}$. For more details on the FUV and
F336W images, we refer the reader to K02.

The Chandra data on 47~Tuc have been described in detail by G01 and
H05. There are actually two distinct data sets, one taken with the ACIS-I
detector in 2000, the other with the ACIS-S detector in 2002. The 2002
observations are considerably deeper than the 2000 ones, so we base
our analysis on the 2002 data set unless specifically noted
otherwise. The ACIS-S plate scale is about 0.5\arcsec~pix$^{-1}$.

In order to astrometrically match up FUV and X-ray sources, we first
applied distortion and boresight corrections to the FUV (and F336W)
data. Our derivation of the boresight correction was based on 11
secure matches (taken from Edmonds et al. 2003a [hereafter E03a]) and
indicates a residual transformation error between HST/STIS and Chandra
positions of only $\sigma_{trans}(\alpha) \simeq 
0.009$~\arcsec and $\sigma_{trans}(\delta) \simeq 0.040$\arcsec. This
level of precision is similar to that achieved by E03a in linking
HST/WFPC2 and Chandra positions. These transformation errors were
added in quadrature to the Chandra positional errors when matching up
X-ray and FUV sources.

Figure~1 shows close-ups of the FUV and F336W images centered on the
Chandra source W31. BSS-7 is seen to be well within the 3-$\sigma$\
error ellipse. In fact, the offset between HST and Chandra positions
is only $\Delta\alpha = 0.060$~\arcsec and 
$\Delta\delta \simeq 0.010$\arcsec, compared to the net (Chandra +
transformation) error of $\sigma_{net}(\alpha) \simeq 0.042$~\arcsec
and $\sigma_{net}(\delta) \simeq 0.064$\arcsec. Thus the FUV source is
located within 1.44-$\sigma$ of the X-ray position. 

Given the radial offset of $\Delta r = 0.061$\arcsec\ between the
Chandra and HST position, we can estimate the probability of a chance 
coincidence between an X-ray source and a BSS in our field of view.
The FUV image is composed of roughly 1 million pixels and includes 19
BSSs and 40 X-ray sources. The positional offset between BSS-7 and
W31 corresponds to 2.09 FUV pixels, so the probability of a chance
coincidence between a BSS and a Chandra source is at most 19 x 40 x
2.09$^2$\ x ${\pi}$ x 10$^{-6}$\ = 0.01. This is actually a slight
overestimate, since roughly half of the X-ray sources turn out to have
genuine FUV counterparts and should therefore be excluded from the
random matching analysis. Thus the chance coincidence probability is 
less than 1\%.

We note that Ferraro et al. (2001) previously suggested a possible
link between W31 and BSS-7, a claim that was subsequently rejected by 
E03a. However, both of these analyses relied on the 2000 Chandra data  
set, in which W31 contributed only 7 counts (G01; H05). In the 2002
data, W31 produced 84 counts (H05), so the astrometry presented here
should be considerably more secure. In particular, the alternative counterpart
for W31 suggested in E03a based on the earlier data is no longer
viable. We stress, however, that the different conclusions arrived at
here and in E03a are mainly due to a 0.43\arcsec\ discrepancy
(corresponding to 4.6${\sigma}$) between the position quoted in G01
for W31 (based on the 2000 Chandra data set) and the position given in
H05 (based on the 2002 Chandra observations). The reanalysis of the
2000 data in H05 reduces this disagreement slightly (to 0.32\arcsec\ or
3.4${\sigma}$). We should finally also acknowledge the possibility that two
different (variable) sources were detected and labelled W31 in the two
Chandra data sets. However, for the purpose of the present study, the
key result is simply the strong match between BSS-7 and W31 in the
2002 Chandra data.

\subsection{HST Time Series}
\label{var}

Given the apparent association between BSS-7 and W31, we decided to
take a closer look at BSS-7 in the extensive HST data set described by 
Gilliland et al. (2000). BSS-7 is saturated in these observations, so
V and I time-series photometry was extracted using the methods of
Gilliland (1994), tweaked to optimize the signal-to-noise ratio for this
particular source. 

The results for the I-band data are shown in Figure~2. A clear
signal with $P = 1.56$~d and amplitude $A \simeq  0.0035$ in relative
flux units (corresponding to 0.0037 mag) is clearly detected. No
corresponding signal was found in the V-band light curve, but the 
photometric scatter in this filter was considerably larger than in
I. In order to test if the non-detection of the signal in V is
significant, we injected a sinusoid of similar amplitude and period as
that detected in the I 
band into the V-band data stream. This confirmed that the V-band data
is too noisy to allow detection of such a weak modulation. The I-band
power spectrum also shows a peak around 97~mins, but this is probably
due to instrumental effects that are modulated on HST's orbital
period (e.g. Edmonds et al. 2003b [hereafter E03b]).

\subsection{Spectral Energy Distribution and Stellar Parameters}
\label{sed}

In order to gain additional insight into the nature of the BSS-7
system, we constructed the broad-band SED shown in Figure~3. This was
achieved by extracting a set of ACS images of 47 Tuc from the HST
archive, all obtained with the High Resolution Camera (HRC). These
images sample the cluster core at a rate of 0.025\arcsec~pix$^{-1}$ and were taken
through a set of filters spanning the full near-UV through far-red 
wavelength range (F250W, F330W, F435W, F475W, F555W, F606W, F625W,
F775W, F814W and F850LP). We then carried out aperture photometry on BSS-7
in all of these images, using a 6 pixel aperture radius. The resulting
measurements were placed on the STMAG system by applying the
appropriate aperture corrections and zero 
points (Sirianni et al. 2005). By deriving a complete set of
magnitudes for BSS-7 from data taken with a single instrument at the
highest spatial resolution, we hope to avoid the systematic errors
associated with simply compiling photometry from the literature. The
HST/ACS magnitudes were finally supplemented with the FUV
(STIS/F25QTZ) magnitude for BSS-7 from K02. This was corrected by 0.08
mag to account for the sensitivity loss of the detector at the time of
the observations (an estimate for this  correction was not available
at the time of K02's publication).

We then proceeded to fit this SED with single and binary models. Our
model SEDs were constructed by carrying out synthetic photometry for
all our filters with the SYNPHOT package within IRAF/STSDAS. 
In doing so, we adopted d = 4510 pc, [Fe/H] = -0.83 and E(B-V) = 0.032
as the relevant cluster parameters (VandenBerg 2000). In all models,
the BSS spectrum was chosen from a grid of synthetic spectra spanning a
wide range of effective temperatures and surface gravities. In binary
models, the secondary parameters were constrained to lie on a 12 Gyr
isochrone (VandenBerg 2000); this is an important constraint that
needs to be kept in mind when interpreting the results. All synthetic
spectra were obtained by interpolating on the Kurucz (1993) grid of
model stellar atmospheres, as implemented in SYNPHOT.

A least squares fit to the SED with a single star model yields BSS
parameters of $T_{eff}(BSS) = 7260^{+15}_{-40}$~K, $\log{g}(BSS) =
3.50^{+0.15}_{-0.15}$ and $R(BSS) = 1.73^{+0.02}_{-0.02} \, R_{\odot}$.
A binary model allowing for a BSS primary and a companion constrained
to lie on a 12 Gyr cluster isochrone (Vandenberg 2000) produces
similar BSS parameters, $T_{eff}(BSS) = 7300^{+50}_{-40}$~K,
$\log{g}(BSS) = 3.60^{+0.15}_{-0.15}$ and $R(BSS) = 1.61^{+0.09}_{-0.11}
\, R_{\odot}$. However, the binary model also points to a
companion on 47 Tuc's upper main sequence, with parameters $T_{eff}(2)
= 5850^{+160}_{-930}$~K, $\log{g}(2) = 4.4^{+0.2}_{-0.2}$, $M(2) =
0.81^{+0.07}_{-0.15} \, M_{\odot}$ and $R(2) = 0.92^{+0.26}_{-0.28} \,
R_{\odot}$. The addition of a binary companion reduced the scatter
around the fit from 0.048 mag to 0.033 mag. According to an F-test,
this improvement is significant at P = 0.077, i.e. about 92\%
confidence. Fig. 3 shows the best-fitting binary model superposed on
the SED. The effective temperature of the BSS suggests a spectral type
late A or early F, and the secondary contributes roughly 15\% of the
flux in the three reddest band passes.

As discussed in Section~\ref{discuss}, the BSS surface gravity
suggested by the SED fits is surprisingly low. We have therefore
checked the sensitivity of the other fit parameters to the assumed 
gravity. For this purpose, we fixed $\log{g}(BSS)$\ at a value
of 4.2 (based on the discussion in Section~\ref{discuss}) and then
refit the data with both single and binary models. For the single BSS
model, this yielded best-fit values of $T_{eff}(BSS) = 7200$~K and 
$R(BSS) = 1.75  R_{\odot}$. For the binary model, the best-fit
parameters were $T_{eff}(BSS) = 7380$~K, $R(BSS) = 1.44 R_{\odot}$,
$T_{eff}(2) = 6000$~K, $\log{g}(2) = 4.17$, $M(2) = 0.87 M_{\odot}$
and $R(2) = 1.3 R_{\odot}$. For the purpose of our discussion below,
these parameters are not substantially different to those derived from
the original fit.

\section{Discussion}
\label{discuss}

In the following sections, we will first attempt to construct a
consistent physical picture of the BSS-7/W31 binary system and then
explore its likely formation, present state, and future
evolution. However, we note from the outset that there are two
possible scenarios for the formation and present state of the system,
which are difficult to distinguish with the existing data. In our
favoured scenario, BSS-7 is a detached binary system with an
upper-main-sequence secondary, with the stellar parameters suggested
by our SED model fits. This scenario implies that at least 3
progenitors are required to form the BSS-7 system. In the alternative
scenario, BSS-7 would be a system that has evolved via mass transfer
from a (possibly primordial) binary. In this scenario, the system is
probably in a semi-detached state at the moment, with the BSS gaining
mass from a low-mass companion that has been stripped of most of its
envelope. In what follows, we will first present the empirical
evidence that BSS-7 consists of a BSS primary with an active
companion, and then explore the two alternative scenarios outlined
above.

\subsection{BSS-7: A Blue Straggler with an Active Companion}
\label{discuss1}

The properties of BSS-7/W31 are unusual. In particular, we are aware
of only five X-ray sources in GCs with candidate BSS counterparts;
these are discussed in more detail in the Appendix. Amongst BSSs
found in open clusters, four have been matched to X-ray sources (Belloni \&
Verbunt 1996; Belloni \& Tagliaferri 1998; Belloni Verbunt \& Mathieu
1998; van den Berg et al. 2004). All are thought to be binary
systems. The lack of X-ray emission from single BSSs in old clusters
is expected, since A- or F-type stars have at most thin convection
zones and thus produce little coronal activity.

Optical variability is more common amongst BSSs in globulars, but
generally falls into two clear categories. First, some BSSs are contact
(W UMa) or semi-detached binaries (Rucinski 2000; A01). These systems
can appear as BSSs due to mass transfer, but most such systems in GCs
have orbital periods P $\ltappeq$\ 0.5 days.
\footnote{However, we note here already that there is at
least one other, probably semi-detached, BSS binary with $P \simeq
1.4$~d in a GC; this will be discussed more carefully in
Section~\ref{alternative}.} Second, some BSSs are
located in the instability strip and are therefore detected as SX Phe-type
pulsational variables. However, these pulsations have even shorter
periods, P $\ltappeq$\ 0.1 days (Gilliland et al. 1998). Clearly, the
optical variability of BSS-7 does not fall into either of these 
categories.

On the other hand, both the period of the optical variations and the
X-ray properties of W31 are typical of the active binary population
in 47 Tuc (H05; A01). These systems are detached, but
still relatively short-period binaries with main-sequence or sub-giant
components. The high level of activity they display is thought to be
linked to the relatively fast spin rate of their active components,
which are usually synchronized with the binary orbit.

Thus all of the evidence we have collected is consistent with the
following physical picture. BSS-7 is a 1.56-day binary system
containing a BSS primary and an active secondary. The 
binary separation must then be $a \simeq 7.1 \left(\frac{M_{BSS} +
M_2}{2 M_{\odot}}\right)^{1/3} \, R_{\odot}$. 

The radius of the BSS primary suggested by our SED fits is $R(BSS)
\simeq 1.6 - 1.7 R_{\odot}$. This is significantly larger than the
radius of a zero-age main sequence (ZAMS) star with the same effective
temperature ($R_{ZAMS} \simeq 1.1 R_{\odot}$). Thus the BSS is already
slightly evolved, which is consistent with the location of BSS-7 in a
FUV-F336W colour-magnitude diagram (K02).

The SED-based surface gravity estimate for the BSS is surprisingly low
for both single and binary models. Taken at face value, this
combination of surface gravity and stellar radius would imply an
unphysically low BSS mass of $M_{BSS} = 0.34^{+0.15}_{-0.08}
M_{\odot}$\ or $M_{BSS} = 0.37^{+0.15}_{-0.10} M_{\odot}$\ for single
and binary models, respectively. 

We can obtain an alternative mass estimate by noting that
sub-giant evolution takes place at roughly constant luminosity. The
mass of the BSS should therefore correspond roughly to that of a ZAMS
star with the same luminosity as the BSS. This yields a more reasonable
estimate of $M_{BSS} \simeq 1.5 M_{\odot}$. We would
then expect our sub-giant BSS to have a surface gravity of $\log{g}(BSS)
\simeq 4.2$, considerably higher than the SED-based estimates.

Experience tells us that the systematic uncertainties on our
photometric gravity estimates may well exceed their formal errors (due
to uncertainties in the cluster distance, reddening and metallicity,
for example). However, it is worth noting that one way to reduce the
effective surface gravity of a star is via rapid rotation. This could
be a direct signature of recent BSS formation in a binary merger (Rasio
\& Shapiro 1995) or an off-axis collision (Lombardi, Rasio \& Shapiro
1996). If this is 
the origin of the discrepancy between expected and measured gravities
for BSS-7, the BSS would have to be rotating near break-up. Taking
$M_{BSS} \simeq 1.5\, M_{\odot}$\ and an equatorial radius of $R_{BSS,eq}
\simeq 1.6\, R_{\odot}$, synchronous rotation with $P = 1.56$\
days corresponds to an equatorial speed of $v_{eq} \simeq
50$\ km~s$^{-1}$; $v_{eq} \simeq 360$\ km~s$^{-1}$\ is required to 
produce an equatorial surface gravity of $\log{g} \simeq 3.6$. The
corresponding  break-up speed is $v_{eq} \simeq
415$\ km~s$^{-1}$. Optical and/or infrared spectroscopy is highly
desirable, both to confirm the composite nature of the SED and to
check for evidence of low gravity and/or rapid rotation of the BSS.

\subsection{The Formation, Present State and Evolution of BSS-7}
\label{discuss2}

\subsubsection{A Detached Binary with an Upper-Main-Sequence Secondary?}   

We now explore our favoured physical picture for BSS-7. In this
picture, the binary is presently detached, and the secondary is an
ordinary upper-main-sequence star. In this case, we can take our SED
fit parameters as valid estimates for both binary companions. 

Certainly the binary separation is large enough to accommodate both
the 1.6 R$_{\odot}$\ BSS and its 0.9 R$_{\odot}$\ companion in a
detached configuration. The observed X-ray emission and optical
variability are also consistent with this scenario. Given that the
companion contributes 15\% in the I-band, the observed variability
amplitude of 0.35\% corresponds to an intrinsic amplitude of about 2\%.
This is in line with the variability amplitudes of other active
main-sequence binaries with similar periods in 47 Tuc (A01).

In the context of the detached binary picture, the formation of the
BSS-7 binary system must have involved at least three cluster
members. The only other observational evidence along these lines in
the literature was based on spectroscopic  mass estimates for five 
BSSs in NGC 6397 that exceeded twice the turn-off mass of their host
clusters (Saffer et al. 2002). However, a recent reanalysis of these
data shows that the \ spectroscopic mass estimates are subject to
rather large uncertainties, and that the earlier estimates were
probably biased too high (De Marco et al. 2005). 

The importance of finding a BSS with 3 or more progenitors derives
from the fact that the two simplest BSS formation channels -- a single
2-body collision or the coalescence of a primordial binary system -- cannot
account for such a system. However, somewhat surprisingly, it is
difficult to rule out evolution directly from a primordial
hierarchical triple for BSS-7 (Iben \& Tutukov 1999). In this scenario, the
inner system would need to have consisted of two roughly equal-mass
components with $M \simeq 0.75 M_{\odot}$\ in order to produce the 1.5
$M_{\odot}$\ BSS we observe today. A binary containing two
such MS stars would need to be near contact to fit stably inside a
1.56-day outer orbit (Kiseleva, Eggleton \& Anosova 1994). This may
seem unlikely, but would be consistent with the fact that this binary
then coalesced to produce the BSS. It should also be kept in mind that
the presently observed 1.56-day period may have been affected by
encounters with other clusters members.

However, still in the context of the detached binary picture, it seems
much more likely that the BSS-7 binary system was formed directly via
dynamical stellar encounters (see, for example, Hurley \& 
Shara 2002; Hurley et al. 2005). One such route is via a single 3-or
4-body resonant interaction. During such an encounter, two (or even
three) stars might have collided to 
form the BSS, while remaining bound to another star involved in the
interaction. The biggest potential problem here is the short orbital
period of the observed binary system. Numerical simulations suggest
that BSS binaries produced via this channel are likely to be much wider
than the binaries involved in the encounter (Davies 1995; Fregeau et
al. 2004). 

Another class of dynamical formation channels invokes (at least) two
separate events to first form the BSS and then the binary system. Thus
the BSS might have captured its companion in a dynamical encounter, such
as a 2-body tidal capture or a 3-body exchange interaction. Once
again, the biggest challenge (at least for an exchange encounter) is
the need to account for the short period of the binary. Most exchange
encounters in dynamical simulations result in binaries with much longer
periods (Fregeau et al. 2004; J. Hurley, private communication).

We finally comment briefly on the future of BSS-7 in the detached
binary  picture. As noted above, the BSS primary has already evolved
off the MS and is currently expanding on 
the sub-giant branch. This must almost inevitably lead to
Roche-lobe overflow. However, since the donor in this case will be
more massive than the accretor, the resulting mass transfer will be
dynamically or thermally unstable (depending on whether the sub-giant
BSS has developed a convective envelope at the onset of mass transfer).
If the mass-transfer ever becomes dynamically unstable, the system
will probably enter a common envelope phase from which an even tighter
binary system might emerge. However, if dynamical instability can be
avoided, the mass transfer may be slow enough to allow the companion
to accrete substantially from the BSS. In this case, the system may
transform itself into yet another BSS binary, with primary and
secondary stars exchanging roles.

\subsubsection{An Alternative Scenario: Semi-Detached Binary Evolution}
\label{alternative}

After this paper had already been accepted for publication, we became
aware of the eclipsing, 1.4~day BSS binary NJL 5 in $\Omega$~Cen (Niss,
J\o rgensen \& Laustsen al. 1978). Given the similarity of the periods
of NJL 5 and BSS-7, it is clearly of interest to ask what is known
about the evolutionary state of the former system. Helt et al. (1993)
have proposed a semi-detached model for NJL 5, in which the primary
and mass gainer is a massive ($M_{1} = 1.2  M_{\odot}$) BSS, and the
donor is a low-mass ($M_{2} = 0.16 M_{\odot}$) star. They also sketch an
evolutionary scenario for this type of system, which starts
with two nearly equal-mass ($M_1 \simeq M_2 \simeq 0.8 M_{\odot}$\
binary components. Such a system can evolve to the present
configuration of NJL 5 via mass-transfer, which is initiated when the
initially more massive star (the current low-mass mass donor) first
fills its Roche lobe. According to this scenario, the donor today
consists of a degenerate Helium core of mass $M_{He} \simeq 0.13
M_{\odot}$, covered by an extended ($R_2 \simeq 1.27 R_{\odot}$)
Hydrogen layer of mass $M_H \simeq 0.03 M_{\odot}$. For details on
this binary evolution path, the reader is referred to Helt et
al. (1993). The future of such a system will probably involve another
phase of (unstable) mass transfer when the BSS expands to fill its
Roche lobe.

At first sight, it may seem that we can exclude this scenario for
BSS-7, since our binary model fit to the SED indicates a secondary on
the upper main sequence. However, it is important to remember that our
binary SED model {\em assumes} that the secondary is on the normal cluster
isochrone. This assumption is incorrect for the low-mass secondary in
the semi-detached model. In our view, an exploration of the full
parameter space of the secondary ($T_{eff}(2)$, 
$log_g(2)$, $R(2)$) in the SED fits is not warranted at present, since
there are only 11 photometric data points which are already quite well
fit by our simpler single and binary models. Thus we cannot rule
out the semi-detached model at present. 

On balance, we nevertheless clearly favour the detached scenario for
BSS-7, for several reasons. First, the location of BSS-7 in the
cluster core makes it likely that the system should have undergone
some dynamical 
interactions in the past (by contrast, NJL 5 is located at about 6
core radii from the center of $\Omega$\ Cen). Second, the X-ray and optical
variability properties of BSS 7 are completely consistent with those
of ``normal'' main-sequence active binaries in 47~Tuc. Third, it
actually seems rather hard to avoid complete binary coalescence in the
Helt et al. scenario. This is because main sequence stars become
deeply convective at masses of about 0.7 and lower, and mass-transfer
from such a star is thought to proceed on a dynamical timescale if the
binary mass ratio is greater than 0.7. This would lead to
coalescence. It is hard to see how BSS-7 could have avoided this fate
during its semi-detached evolution. It is worth noting that this last
comment also applies to NJL 5, and that the light curves analysed by
Helt et al. (1993) actually cannot distinguish between semi-detached
and detached models for this systems. However, Helt et al point out
that Margon \& Cannon (1989) have derived an upper limit of
30~km~s$^{-1}$\ for the radial velocity of the BSS primary in NJL
5. This would suggest that the secondary in NJL 5, at least, may
indeed be a low mass star. 

In any case, the cleanest proof of the 3-progenitor model for BSS-7
would be to establish conclusively that the total system mass exceeds
twice the cluster turn-off mass (as is indeed the case for our
best-fitting binary SED model). The system parameters needed to carry
out such a test should be obtainable via time-resolved spectroscopy of
the system.

\section{Conclusions}

We have shown that the bright blue straggler BSS-7 in the core of
47~Tucanae is the counterpart to the Chandra X-ray source W31. We have
also analysed optical light curves of BSS-7 and find a clear 1.56~day
periodic signal in the I-band. We finally constructed a broad-band FUV
through red SED for the system and fit this with single and binary
models that take the secondary to lie on the cluster isochrone. 

All of our results are consistent with the interpretation that
W31/BSS-7 consists of a massive BSS primary with an active, upper-main
sequence companion. If this is correct, then at least 3 progenitors
are needed to form this binary system. This would rule out the simplest
models for BSS formation -- binary coalescence or a single physical
collision -- but would be consistent with N-body models in which blue
stragglers often form via multiple encounters that can involve both
single and binary stars. 

However, even though we favour the scenario outlined above, we cannot
completely rule out the possibility that BSS-7 has 
descended directly from a binary system via mass transfer. In this
case, the secondary is a low-mass star with a Helium core that has
been stripped of most of its Hydrogen envelope. Time-resolved
spectroscopy is strongly encouraged in order to distinguish
definitively between these scenarios.

\acknowledgments

We are grateful to Tom Maccarone for useful discussions and to Jarrod
Hurley for his help in understanding the possible formation and
evolution channels for BSS-7 and NJL 5. We also thank Peter Edmonds
and Craig Heinke for pointing out the existence of other X-ray/BSS
matches in 47~Tuc.

\appendix

\section{Blue Stragglers in Globular Clusters with Candidate X-ray
Counterparts (and some Related Sources)}

We are aware of only five X-ray sources in GCs with candidate BSS
counterparts. These are W92, W163, W167 and W266 in 47~Tuc (E03; H05)
and XMM source \#20 in M22 (Webb et al. 2004). 

W92 is the most interesting of these. It has an X-ray luminosity of
L$_X$(0.5-6.0 keV) = $2.3 \times 10^{30}$~erg~s$^{-1}$\ in the 2002
Chandra data (H05), is located at about 2 core radii from the cluster
center and was identified by E03 with the 1.34-d eclipsing binary
WF2-V03 discovered by A01. The astrometric match between W92 and
WF2-V03 is good to 0.07\arcsec, which is comparable to that between
W31 and BSS-7. If the identification W92 = WF2-V03 is correct, this
BSS binary may be very similar to W31 = BSS-7. The fact that WF2-V03
is eclipsing would make it particularly valuable for follow-up 
observations aimed at measuring its system parameters.

W163 and W167 in 47~Tuc are also located outside the cluster core and
have been matched to optically variable BSSs by E03 (the former match
is to source WF3-V05 in A01). However, the optical periods of the
counterparts in these cases are between 0.3 and 0.4 days, so they are
probably W UMa binaries. These systems may appear as BSSs because of
mass transfer from one binary companion to another, so there is no
need to invoke three progenitors to account for these binary BSSs. 

The optical counterpart suggested by H05 for W266 in 47~Tuc,
source PC1-V10 from A01, is also a W~UMa star (A01). However, W266 is
located inside the cluster core and within the FUV field of view. In
this case, our own astrometric analysis suggests that the probability of a chance 
coincidence is about 12 times larger than for W31, and that at least
two FUV sources lie closer to W266 than the BSS. These results will be 
discussed in more detail in a separate publication. 

XMM-20 in M22 (Webb et al. 2004) is located just outside the half-mass
radius of this cluster. The  positional match between this X-ray
source and the suggested optical counterpart is only good to
4.2\arcsec, but given the lower spatial resolution of XMM and the
relatively low stellar density in the periphery of M22, this is
nevertheless a viable match with a small probability of chance
coincidence. 

All of these sources, but particularly W92 in 47~Tuc, deserve closer
observational scrutiny to confirm the reality of the X-ray/optical
matches and to determine the nature of these systems. 

We finally point out 3 other BSS that have not been suggested as X-ray
counterparts, but that are nevertheless particularly
interesting. These are (i) PC1-V12 in 47 Tuc, which was found by A01
to be an optically variable BSS and possible active 
binary; (ii) V18 in NGC6397, which Kaluzny \& Thompson (2003) classify
as a likely detached eclipsing binary with a period of 0.8~days; (iii)
V2 in Terzan 5, which was suggested by Edmonds et al. (2001) to be a
likely eclipsing BSS with $P_{orb} \simeq 14$~hrs. These sources, too,
deserve closer observational scrutiny.


\clearpage

\figcaption[f1]{Far-ultraviolet (left panel) and optical (right
panel) close-ups on the vicinity of W31. The white ellipses mark the
3${\sigma}$ error region around the X-ray positions in the 2002
Chandra data set. Black circles edged in white mark the location of
BSS-7, our proposed counterpart for W31. Both images are roughly
1\arcsec $\times$\ 1\arcsec, with the long axes of the ellipses being
oriented north-south. More specifically, north is up and to the left
and east is perpendicular to this axis and down.}

\figcaption[f2]{I-band light curve (top panel), power spectrum (middle
panel) and folded light curve (bottom panel) for BSS-7 in the HST
data set of Gilliland et al. (2000). A 1.56-day periodic signal is
clearly visible.}

\figcaption[f2]{The broad-band spectral energy distribution of
BSS-7. All photometric measurements are plotted STMAG magnitudes at
the mean wavelength of the relevant band pass. The 
STMAG system is defined as $m_{STMAG} = -2.5 \log{F_{\lambda}} -
21.1$, where $F_{\lambda}$\ is the flux of a flat
spectrum source that would produce the observed count rate. Also shown
are synthetic spectra for the best-fitting binary model (primary,
secondary and combined binary). Note that the model spectra are shown
as continuous curves solely for illustration; they were folded through
the appropriate filter transmission curves for the purpose of fitting
the SED. The bottom panel shows the residuals of the least-squares fit. 
The error bars shown are those needed for the best-fitting model to
achieve $\chi^2 = \nu$, where $\nu = N_{data} - N_{par}$\ is the number
of degrees of freedom. In binary model fits to the 
SED there are $N_{data} = 11$\ data points and $N_{par} = 4$\ free
parameters (effective temperature, gravity and radius for the primary,
plus the location of the secondary on the cluster isochrone).}

\clearpage

\begin{figure}[t]
\plotone{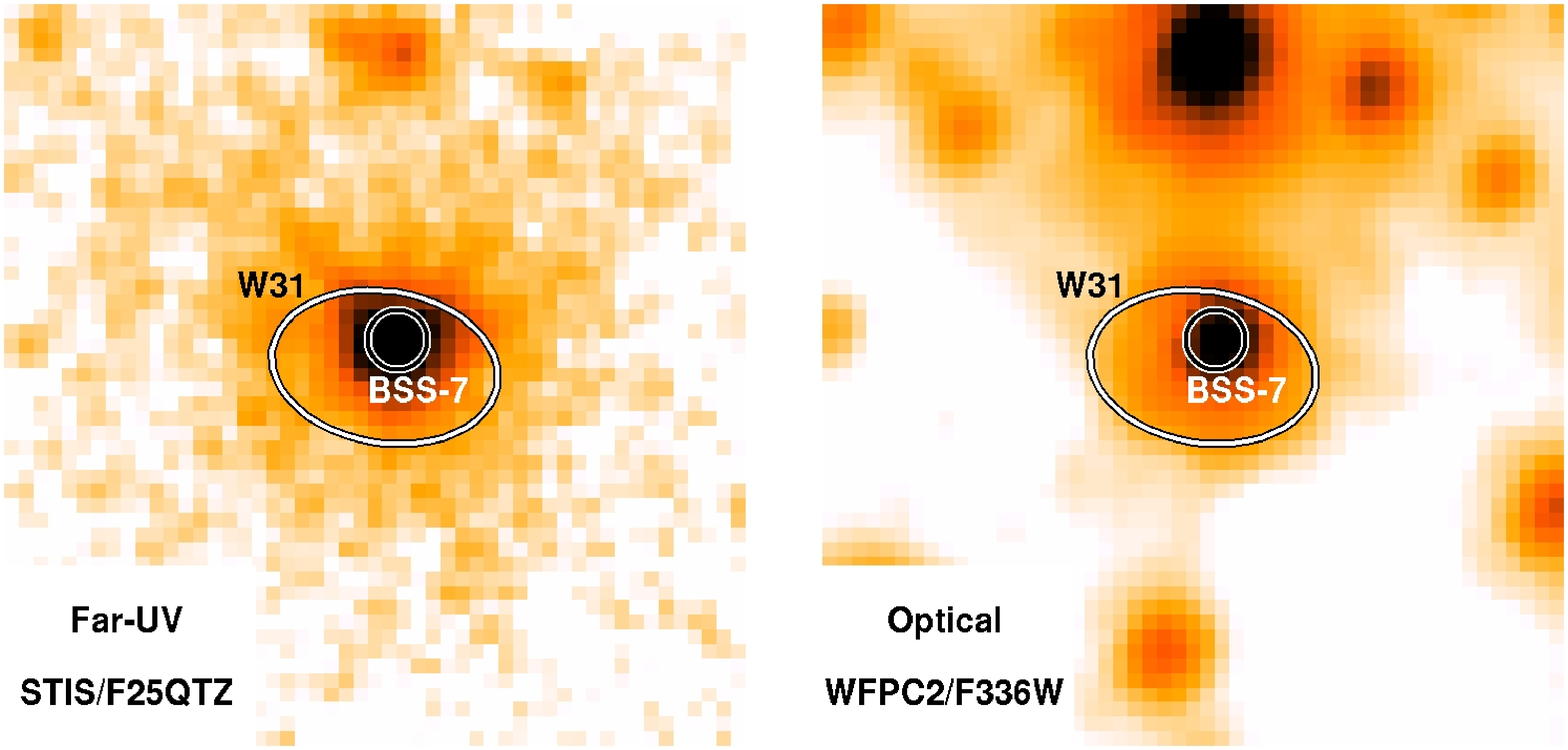}
\label{f1}
\end{figure}

\clearpage

\begin{figure}[t]
\plotone{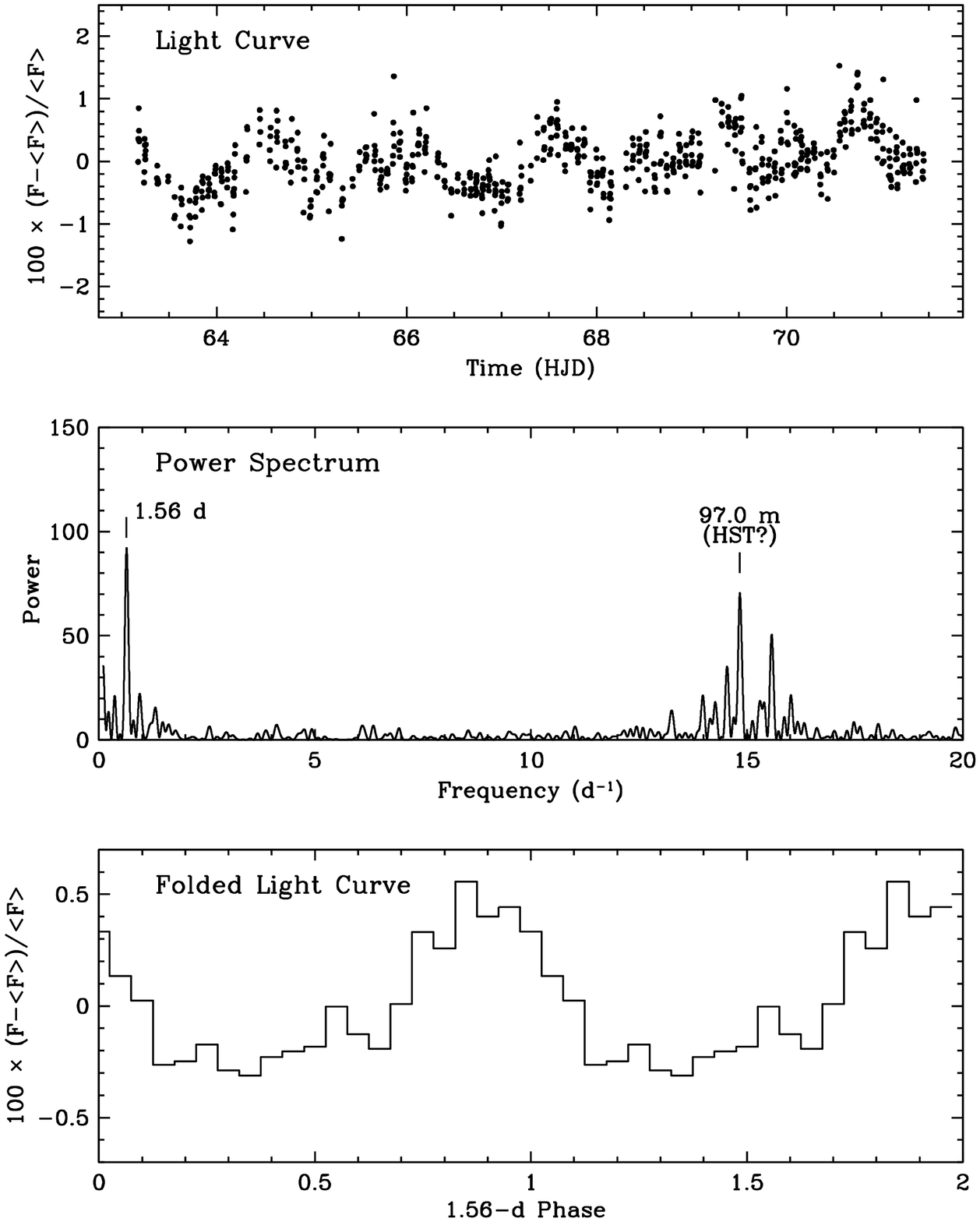}
\label{f2}
\end{figure}

\clearpage

\begin{figure}[t]
\plotone{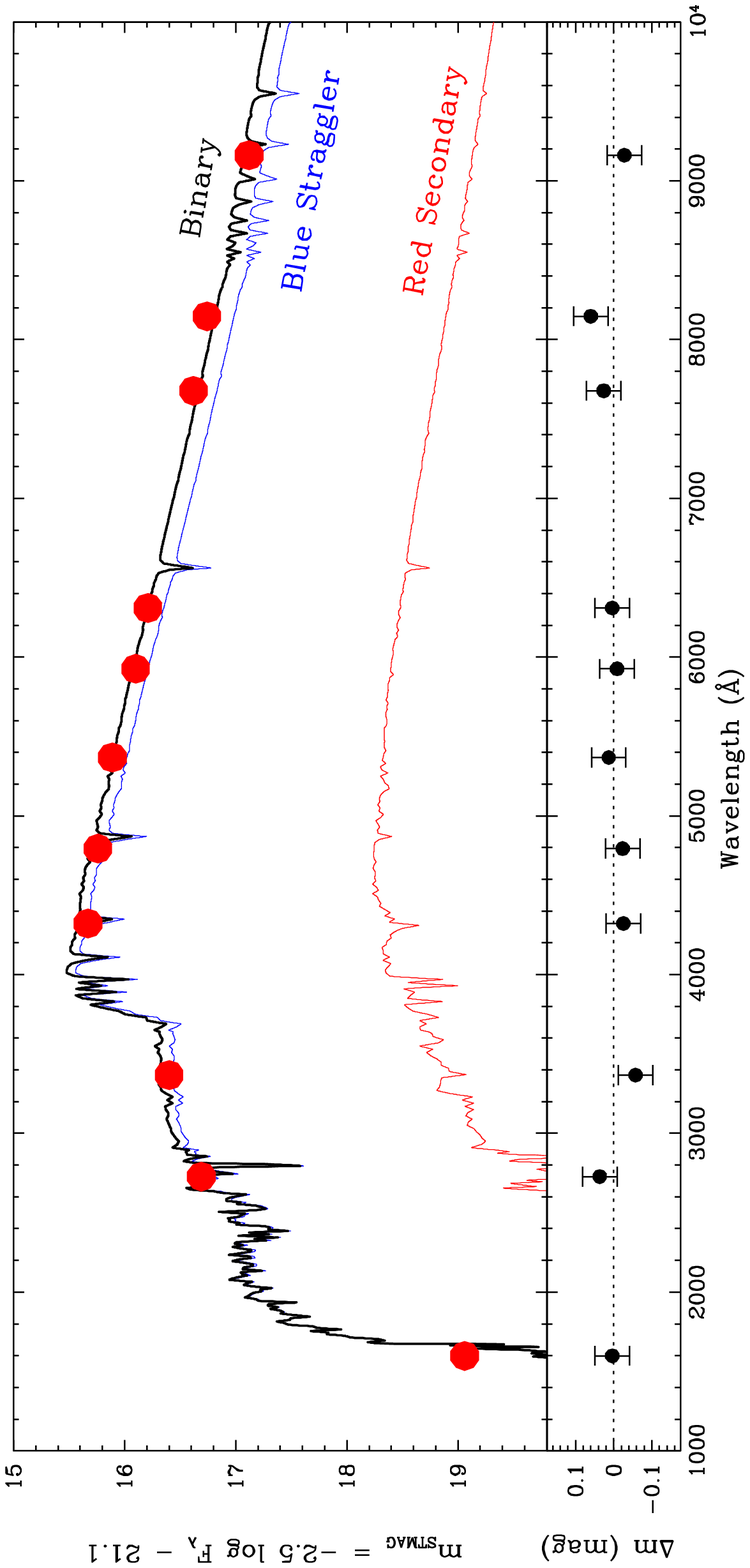}
\label{f3}
\end{figure}


\end{document}